\documentclass{article}

\usepackage{url}
\usepackage[pdftex]{graphicx}

\begin{document}

\title{A survey on social network sites' functional features\thanks{Preprint of article published in Proceedings of IADIS WWW/Internet 2012. Madrid, Spain}}

\author{Antonio Tapiador and Diego Carrera \\
\small{Universidad Polit\'{e}cnica de Madrid} \\
\small{\{atapiador, dcarrera\}@dit.upm.es}
}

\date{}

\maketitle

\begin{abstract} 

Through social network sites (SNS) are between the most popular sites in the Web, there is not a formal study on their functional features. This paper introduces a comprehensive list of them. Then, it shows how these features are supported by top 16 social network platforms. Results show some universal features, such as comments support, public sharing of contents, system notifications and profile pages with avatars. A strong tendency in using external services for authentication and contact recognition has been found, which is quite significant in top SNS. Most popular content types include text, pictures and video. The home page is the site for publishing content and following activities, whilst profile pages mainly include owner's contacts and content lists.

\end{abstract}

\section{Introduction}



Social network sites (SNS) are between the most important sites in the web \cite{Mislove:2007}. However, there is not a formal survey of their functional features.


McKeever \cite{McKeever:2003} and Liduo \cite{Liduo:2010} provide such a list of features in the field of content management systems (CMS).

There exist recent surveys in the field of social computing. King et al. surveyed computational approaches in social computing \cite{King:2009}. They look at social platforms, such as social networks and other social media. Their work is focused on computational tasks and techniques, such as social analysis, ranking, query log processing, or spam detection, without approaching the field of functional features.

boyd and Ellison \cite{Boyd:2007} define SNSs as web-based services allowing individuals to do three basic tasks, i.e. construct a profile, articulate a lists of users with whom they share connection and view and traverse the lists. They do talk about other elements that SNSs may have: avatars, privacy settings, customization of relation names (beyond the “friend” cliché), posting resources to user’s wall, private messages and any type of content sharing (photos, videos, etc.) However, they do not provide a formal measure on how these features are adopted by current social network platforms.


This article introduces a survey on common functional features of SNS. A thorough list of social network features has made up, along with their description and their utility in a social context. Then, 16 high popular social network platforms have been analyzed, evaluating which features are implemented by them. The paper ends with discussion of the results and some conclusions. 

\section{Enumeration of SNS functionalities}

We have made up a comprehensive list of functional components that can be found in social network sites. Some of them were already present in \cite{Boyd:2007}, while others are novel in the literature.

A social network is a finite number of actors and the relations defined between them. The term was originally attributed to Barnes \cite{Barnes:1954}. The presence of information about relations is a main characteristic of a social network. The concepts behind a social network are actors and their relations \cite{Wasserman:1994}. These are the basic elements that need to be present to start talking about a social network.

\subsection{Social actors}

Social entities are referred to as actors. They can be people of course, but also collective social units such as corporations, organizations, institutes, cities or even events. The term actor does not mean that the entity has the ability to act by itself, but it may usually treated in that way, for instance, \textit{"The University has signed an agreement"}, though is the head of the University which in fact performs the action in behalf of the institution.

Social actors must be explicitly supported by a social network platform. This feature should be implemented by every site that claims to be a SNS.

\textbf{User accounts}; users must be allowed to register in the site. They should fill a form, provide some basic data, typically name, email and password, though this could vary from one site to the other. Later, in the log-in process, users provide credentials in order to be authenticated. The importance of authentication resides in that users perform actions in their own behalf, like establishing relations with other users, uploading content, etc. The requirement is that users must be univocally identified. The website also bases authorization on user authentication.

Some SNS may force the \textbf{email to be confirmed} and do not let users perform actions in the site until this step is completed. On the other hand, there are sites that let users continue and start exploring the site without email confirmation.

Nowadays, leading websites are providing authentication services to other websites. SNS may \textbf{delegate authentication} to them by using these services. This is a convenient and user-friendly way to authenticate, because users do not have to introduce their credentials just another time, and the SNS can extract profile data and contacts from the authentication service. 

Some features intend to bring more users to the SNS. One of them is sending \textbf{invitations} to other users by email. This feature consists on a form in which users introduce their contacts' emails. An extended version of the former functionality is \textbf{contact import}. The SNS obtains the agenda from the user's email account or other SNS, and uses it to send invitations or set user's contacts in the site.

\textbf{Other social entities} should be also supported by the site, such as \textit{groups}, \textit{organizations}, \textit{institutions}, etc. This is a way of extending coverage to other entities in the social spectrum. These entities may be able to maintain their own social image, which includes logo, contact information, etc. They must not be confused with contact groups or lists, which are reviewed below.

\subsection{Social relations}

In SNS, actors must be able to establish some kind of relations with other actors. This must be another mandatory feature for every SNS. Relations can be unidirectional o bi-directional.

\textbf{Unidirectional relations} are established by one of the parties and do not need confirmation by the other party. \textit{Follow} is a popular case of unidirectional relation. When some actor \textit{follows} another one, he is expressing his interest in the activity of the other actor. This usually implies that their activities reach them to their home wall, as we will see below, in section \ref{home-wall}. The paradigm of this relation is Twitter. A full functional social network can be build with unidirectional relations \cite{Tapiador:2011:TieRBAC}.

\textbf{Bidirectional relations} mean that both parts need to approve the relation to be established. Facebook is the paradigm of this relation type. Bidirectional relations are usually attached to privacy and permissions; e.g. when both users accept the relation, they are allowed to publish to their walls. However unidirectional relations can also provide them \cite{Tapiador:2011:TieRBAC}.  

The way to establish relations is usually facilitated with a link or button next to the small card of users, next to their avatar, name and other data.

\textbf{Contacts management} features help users organize their contacts in several categories, lists or groups. It is a way to avoid the context collapse problem \cite{Boyd:2007}. Contacts management is in constant evolution and there has been a lot of research related to privacy and permission management. There are several models to manage contacts; Twitter uses lists, as well as Facebook. Google+ uses a more advanced and eye-candy interface of Google Circles. Contact management settings can also be used in \textbf{privacy settings} to grant permissions to certain groups of contacts at the same time.

\subsection{Content}

Contact-oriented SNS usually provide means for their users to share content. Besides, \textit{content-oriented} SNS are built around some specific type of content \cite{Tapiador:2011:ExtendedIdentity}. These features take the influence from the work in web content management systems (CMS) \cite{McKeever:2003}.

\textbf{Content types} managed by SNS may be very different. Text comments are the most basic type of content. Examples of SNS built around content are Flickr for photographs, YouTube for videos, Github for code repositories. Even contact-oriented sites like Facebook support some kinds of contents such as pictures or events. One type of content that is very popular in contact-oriented networks is external links to other websites. Using this feature, contact-oriented sites become a forum for content exchange between affine people or groups.

Specific type of content help to focus the website on some specific type of user, e.g. software repositories or artistic photographs will attract software developers and photographers, respectively. The more types of content the social platform supports, the more channels actors have to interact between them, but also the more diluted the purpose and identity of the SNS becomes.

Content may have \textbf{threads} that allow discussions on topics. Content have a “reply” button that allows submitting new related content, such as \textbf{comments}. Afterwards, new content appears below and often tabulated to the right. Content threads encourage engagement between users.

Another functionality providing engagement is to \textbf{cite or tag other users} in content. This is typical in Twitter, when users are cited in tweets by writing their name preceded by \textit{@}. Other types of citation are facial recognition and photo tagging. The last one is very popular on Facebook, and many people use it as some kind of heads up.

SNS may also implement \textbf{content search engines}, a feature that enables users to find content that was uploaded to the network. This feature also comes from the CMS world.

\subsection{Communication tools}

SNS sites may also support other \textit{communication tools}. One example is \textbf{system notifications}, by which the system alerts users of some types of events, for example the posting of new content that may be of their interest. Other examples include \textbf{private messages} between users, \textbf{chat} and \textbf{video-conferences}. These communication tools are often connected to email.

\subsection{Privacy and content visibility}

Privacy in social networks is a hot research topic nowadays. Access control mechanisms allow content to be shared only with a given audience. When posting content to the site, users can choose which users or groups of people they want to share with.

There are different levels of information disclosure. \textbf{Private} content means that only the user that posted the piece of information to the SNS will have access to it. Other option may be sharing the content only with current \textbf{contacts}. An extended option may be using second degree relations, so sharing the content until \textbf{contacts of contacts}. Audience can also be picked up from groups or lists in \textbf{contact management settings}. For instance, Google+ allows choosing the audience of content from user-defined Google Circles. Another option is sharing with \textbf{specific users}, which may be chosen from existing contacts. Finally, content may be shared \textbf{publicly}.

The existence of different levels of content visibility may facilitate users to share more content. There are some sensitive pieces of information that would not be shared if the only option is publishing them to the public, but they might be shared with a restricted audience.

\subsection{Ratings}

In many social network sites, people are allowed to rate content and activities from other users. \textbf{Rating systems} take several forms, from like, binary like and dislike, 5 starts, etc. Rating systems are reviewed at Sparling and Sen \cite{Sparling:2011}. This functionality is the base for popularity rankings. They can be used for building recommendations to users and promoting featured content.

\subsection{Activities timeline}

An activities timeline consists on a more-recent ordered list of actions performed by one or several actors. It may refer to posting some content, creating new friendships, rating content, etc. Examples of these actions are \textit{"Alice uploaded a photo"}, \textit{"Bob likes your comment"} or \textit{"Charlie pushed three commits to a repository"}.

The activities timeline help users to be aware of the activity of others, and have a sense of what is happening in the social network. They can appear in several parts of the site, as we will see below.

\subsection{Wall}

We define a wall as the posting box that allows actors adding or creating any type of content to his own profile or to other actor in the SNS. The box is usually besides the activity timeline, so after the user posts a new content, a new activity appears in the timeline, a way of providing users with \textit{instant gratification}.

The box usually supports posting several types of contents, such as text and files. It can also appear in several sections of the SNS, as we will see next.

\subsection{Home}

As users are authenticated, SNS may have a special page called home or dashboard. This is the first page offered to users after signing into the site. This page is used to present users a summary of the most important information in the user's social network.

\label{home-wall}

Some features that may appear in the home page include a \textbf{wall}, where the user can post new content to the social platform, an \textbf{activities timeline} with the actions from the people in his network, and relevant \textbf{contact or content lists}.

\subsection{Profile}

Social network sites may offer a single web page per actor registered in the network. These pages are quite important, as they represent the external image that social actors project, they are the representation of users to the rest of the social network \cite{Boyd:2007}. They build a summary of their identity.

Profile pages may have the following sections:

\begin{itemize}

\item \textbf{Profile information} about the actor, such as location, age or birthday, contact details, etc. This may be a way to increase trust, helping users to recognize others and promoting interaction between them.
\item An \textbf{avatar}. Actors can upload and crop an image that represents them in the website. This feature also may help to build trust, in the same way as the former.
\item \textbf{Activities timeline} a list of more-recent ordered actions that includes only activities related to the profile owner. It provides a way to show a summary of recent profile owner's activity in the SNS.
\item \textbf{Wall} allows the profile owner and other actors to post activities to him or her, and thus, improve communication between users.
\item \textbf{Contact list}: a list of the contacts from the profile owner. It helps the network growing, letting the users to browse through the social network and finding more contacts.
\item \textbf{Content list}: a list of relevant pieces of content shared with or by the profile owner.
\end{itemize}

\section{Survey}



We have surveyed 16 top social network platforms. The sites were chosen from Wikipedia's list of social networks websites \footnote{\url{http://en.wikipedia.org/wiki/List_of_social_networking_websites}}, ordered by page ranking. We excluded non-English sites and sites that had restricted sign-up and needed invitation. Table \ref{table:site_list} shows the lists of sites surveyed, along with their description, registered user number and Alexa's page ranking. SNS users are between almost 1 billion from Facebook to 10 million from SoundCloud. Page rankings are between 2nd from Facebook to 435th from Viadeo. SNSs' focuses include general, blogging and micro-blogging, photo sharing and art, business and music sharing.

\begin{table*}[ht]
\caption{List of surveyed SNS, along with their registered users and pange rankings. Source: Wikipedia} 
\centering 
\begin{tabular}{|c|l|c|c|} 
\hline 
Name & Description/Focus & \begin{tabular}[c]{@{}c@{}}Registered \\ users\end{tabular} & \begin{tabular}[c]{@{}c@{}}Alexa's \\ ranking\end{tabular} \\ 
\hline 
Facebook & General & 908,000,000+ & 2 \\ 
\hline 
Twitter & General. Micro-blogging & 500,000,000 & 8 \\ 
\hline 
LinkedIn & Business and professionals & 160,000,000 & 12 \\ 
\hline 
Flickr & Photo sharing & 32,000,000 & 48 \\ 
\hline 
LiveJournal & Blogging & 17,564,977 & 115 \\ 
\hline 
Badoo & \begin{tabular}[l]{@{}l@{}}General \\ Meet new people and dating \end{tabular} & 154,000,000 & 118 \\
\hline 
deviantART & Art community & 22,000,000 & 131 \\
\hline 
StumbleUpon & Stumble through websites & 20,000,000 & 146 \\
\hline 
Myspace & General & 30,000,000+ & 161 \\
\hline 
Yelp, Inc. & Local Business Review and Talk &  & 186 \\
\hline 
Taringa! & General & 11,000,000 & 214 \\
\hline 
XING & Business & 11,100,000 & 270 \\
\hline 
Tagged & General & 100,000,000 & 288 \\
\hline 
SoundCloud & Music pieces & 10,000,000 & 299 \\
\hline 
Orkut & General & 100,000,000 & 319 \\
\hline 
Viadeo & \begin{tabular}[l]{@{}l@{}}Social Networking and \\ Campus Networking\end{tabular} & 35,000,000 & 435 \\
\hline 
\end{tabular}
\label{table:site_list} 
\end{table*}

Two user accounts were opened for the survey, using Gmail and Hotmail email providers. The survey was performed between July 10th and July 14th, 2012, and consisted in the following steps:

\begin{enumerate}

\item Go to site's main page. Does the site allow signing up? Are there any other authentication services available to perform the registration?

\item After signing up providing an email, does the site allow performing actions? Or is email confirmation needed?

\item After signing up with both accounts, try to find the other user. Does the site have actor search? Can invite by email? Can use external agendas (email, other social networks) to search for contacts? If cannot search for the other account, or the search does not work, just paste the profile page from one browser to another.

\item When the other account is found, add him as a contact. Does the site require confirmation from the other part (bidirectional) or not (unidirectional)? Can the contact be blocked? Is there any type of contact management available? Are privacy settings related to contact management settings? 

\item What type of contents can be posted to the site? Can perform comments? Re-share? Tag or mention other users? Search for content? 

\item Does the site provide system notifications? What about private messages, chat or video-conference?

\item When posting content, what type of access control is available? Options are: make the content private, share with all the contacts, with contacts of contacts (second degree), settings from content management, like user lists, share with specific users or make the content public.

\item Can users rate the content? Available options: like, like or dislike or any kind of scale

\item Is there a home page available? Does it include a contact list, content list, activities timeline or wall?

\item Is there a profile page? What kind of actor's attributes does it contain? Does it include an avatar? What about contact lists, content lists, activities timeline or wall (both in own profile and other's profile)?

\item Finally, is there any other type of social entities supported by the site?

\end{enumerate}

\section{Results}

\begin{figure}
\centering
\includegraphics[scale=0.5,trim=0 0 110 0,clip=true]{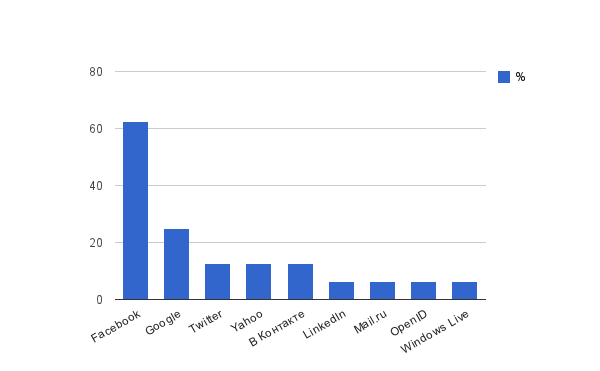}
\caption{Usage of external authentication services}
\label{fig:authentication-services}
\end{figure}

\textbf{Every site allowed users to create an account}. Some of the services provided the option of using \textbf{external authentication services}. Figure \ref{fig:authentication-services} shows the most popular services used by the surveyed social platforms. Facebook is the most popular service for external authentication, used by more than 60\% of the sites analyzed. Google follows, with more than 20\%, then Twitter and Yahoo.

Regarding the \textbf{immediate use of the service}, the 68.75\% of the analyzed platforms allowed to perform operations in the site without confirming the email address. However, some of them restricted the actions, e.g. uploading content or establishing contacts. Others denied subsequent logins without the email account being confirmed.

\begin{figure}
\centering
\includegraphics[scale=0.5,trim=0 0 110 0,clip=true]{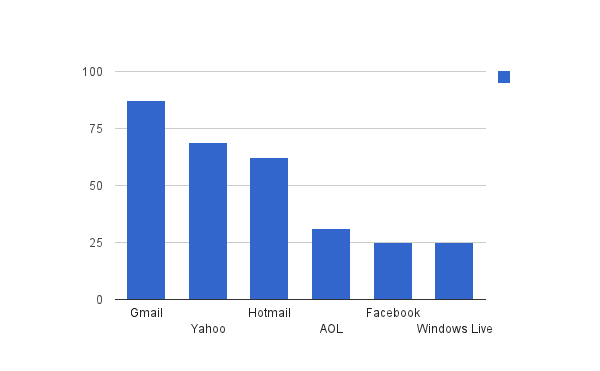}
\caption{Usage of external services to import friends to the SNS}
\label{fig:import-friends}
\end{figure}

Almost one third of the websites (31.25\%) implement the feature of \textbf{invite external contacts} with a form by sending them an email. This figure contrasts with the high availability to \textbf{search in email agendas or other SNS} contacts. Figure \ref{fig:import-friends} shows the support for searching in other services for contacts. Gmail is supported in 87.5\% of the SNS for leveraging contact recognition. Yahoo and Hotmail services follow with 68.75\% and 62.5\% respectively. Note that other SNS such as Facebook is also used for this case in a 25\% of the cases. Almost other 100 different services were supported by analyzed SNS, notably Badoo and LinkedIn support more than 50 each.

Almost all the SNS analyzed supports \textbf{searching for users}, being Taringa! the exception.

\begin{figure}
\centering
\includegraphics[scale=0.5,trim=0 0 110 0,clip=true]{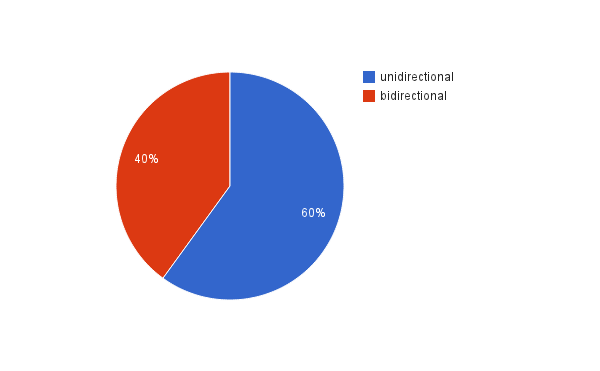}
\caption{Support for relationship types in SNS}
\label{fig:relationships}
\end{figure}

\textbf{Every SNS allows some kind of relationship} between users. Yelp, Inc., was the only one that supports both unidirectional and bidirectional relationships, SNSs tend to support only one relation mode. Figure \ref{fig:relationships} shows the distribution of relation mode between SNS. \textbf{Unidirectional relationships are slightly more used by SNS}.

Some type of \textbf{contact management is supported in half of the cases} (50\%). 25\% of the sites provide their users with preset lists. Regarding permissions, privacy settings can be shared with contacts more than half of the cases (56.25\%), however using specific content management settings for permissions is available in 12.5\% of the SNS.

More than half of the sites (62,5\%) support \textbf{blocking other users} among their features.

\begin{figure}
\centering
\includegraphics[scale=0.5,trim=0 0 110 0,clip=true]{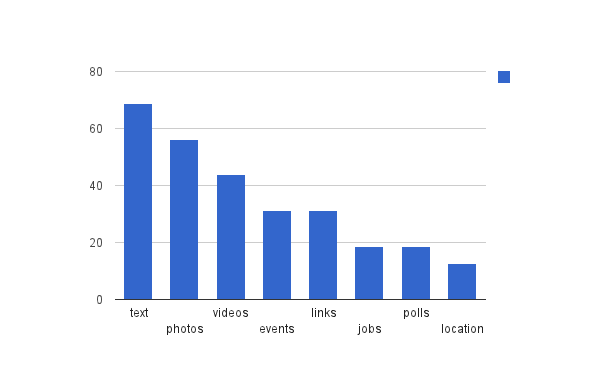}
\caption{Most popular first-post content types managed by SNS}
\label{fig:content-types}
\end{figure}

Figure \ref{fig:content-types} show the most popular \textbf{content types} that are posted first time. It excludes comments. Text posts are the most popular type (68.75\%), followed by photos (56.25\%), videos (43.75\%), events and links (31.25\%). There is a tail of other different type of contents supported by sites, depending of what the SNS is focused on. They include reviews, ratings, prints or games.

Comments do not appear in the graphic as they are not the start of the \textbf{thread}, but they are supported by every SNS. {Reshare} contents from other users is supported by half of the sites (50\%). On the other hand, \textbf{tag or mention users} is supported by more than third of the SNS (37.5\%). \textbf{Content search} is supported by every SNS.

Regarding communication tools, \textbf{system notifications} are present in every SNS. \textbf{Private messages} are supported in 93.75\% of the sites, only not supported by Orkut. \textbf{Chat} is not so popular, supported only by 18.75\% of the sites. Finally, \textbf{video-conference} is not present in any site.

\begin{figure}
\centering
\includegraphics[scale=0.5,trim=0 0 110 0,clip=true]{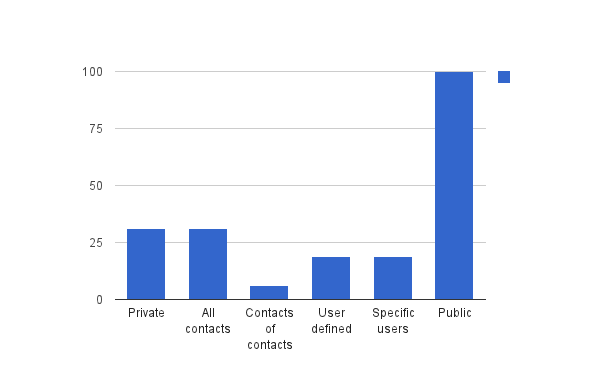}
\caption{ Content privacy settings supported by SNS}
\label{fig:privacy-settings}
\end{figure}

Regarding \textbf{content visibility}, there is an irregular distribution between the available options. Figure \ref{fig:privacy-settings} shows the popularity of options by SNS. Uploading content and keep it private, and sharing only with direct contacts both features are supported by 31.25\% of the sites. Sharing with contacts of contacts is only supported by Facebook. Sharing with user-defined contact lists, or with specific-users is supported by few sites. Finally, \textbf{every SNS supports public sharing}.

Unary \textbf{rating} is the most popular one; it is supported by 68.75\% of the sites. Like and dislike is only supported by one site. There is also only one case with a scale of 5.

\begin{figure}
\centering
\includegraphics[scale=0.5,trim=0 0 110 0,clip=true]{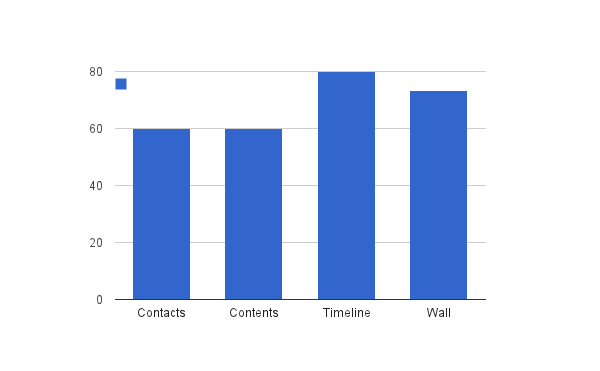}
\caption{Home page features present in SNS}
\label{fig:home-features}
\end{figure}

Regarding the \textbf{home page}, this feature is present in almost every SNS, being devianART the exception. Figure \ref{fig:home-features} shows features present in home pages. A list of, or at least a link to, contacts and content is present in more than half the SNS, while the timeline is much more usual (80\%). Finally, posting content from the homepage is well supported, almost as much as the activities timeline. 

\begin{figure}
\centering
\includegraphics[scale=0.5,trim=0 0 110 0,clip=true]{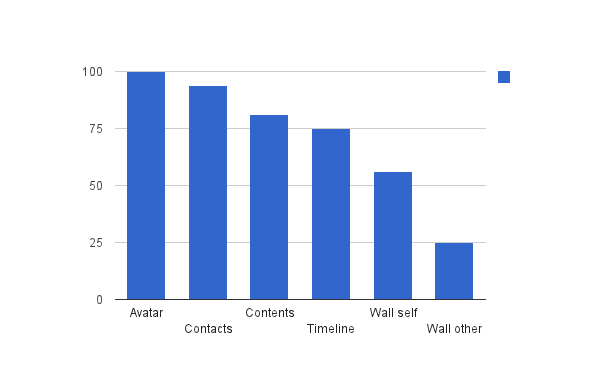}
\caption{Profile page features present in SNS}
\label{fig:profile-features}
\end{figure}

\textbf{Every SNS has a profile page per user}. Figure \ref{fig:profile-features} shows popular features in profile pages. \textbf{Profile attributes} varies much from one site to another, with an average of 2 attributes per profile, and \textit{location} being the most common. \textbf{Avatar} support is outstanding, being present in every site. \textbf{Contacts and contents lists} are also quite present in profile pages. Activity timeline is less popular, but still has a 75\% of present in sites. Posting in self profile is less supported, only by 56\% of the sites. Finally, posting to other users in their profile is only supported by 25\% of the SNS. 

\begin{figure}
\centering
\includegraphics[scale=0.5,trim=0 0 110 0,clip=true]{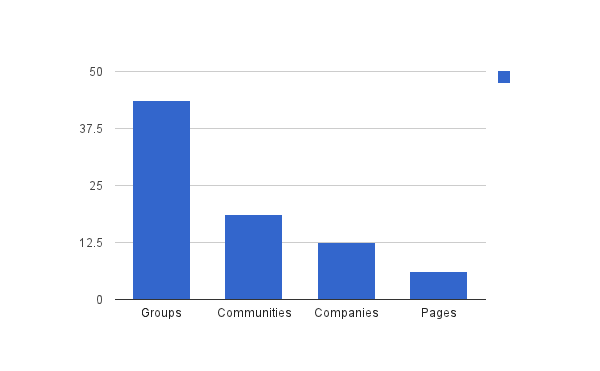}
\caption{Social entities supported by SNS}
\label{fig:social-entities}
\end{figure}

Finally, figure \ref{fig:social-entities} shows other types of social entities supported by SNS. Groups are the most popular social entity, being present in almost half of the sites (43.75\%). Some sites like Facebook and XING support two types of social entities at a time (groups, pages and groups, companies respectively).

\section{Discussion}


Results show a list of functional features supported by every SNS. These features include user account creation and relationship establishment, as it was expected by the social network definition. This is also consistent with the idea that SNS base identity building and network creation on authentication. Other features present in every SNS are comments, some type of content publishing with public visibility, system notifications and a profile page with an avatar. Features present in almost every social network include user search, private messages and a home page. These results contribute to the set of features mentioned by boyd and Ellision \cite{Boyd:2007}.

We have found a very strong tendency to delegate authentication. Being present among the top popular sites, which fight between each other for a higher user base, we expect more usage of this feature in not so popular platforms, which can take more advantage of it to leverage authentication.

Friend recognition using external services is other popular feature, with Gmail support present in 87.5\% of the SNS. Results on Facebook's friend recognition may be higher if authentication were performed using Facebook, instead of using an email-based sign up.

Regarding relation types, unidirectional and bidirectional options are both almost equally popular. Blocking relation is significant, present in more of the half of the cases.

Another significant finding is the presence of comment support in every SNS. Content types have popular ones, such as text, images and videos, and other types are more specialized. We expect these types be more diverse as analyzing other SNS, as our sample is biased towards the most popular, general purpose SNS. We find here an opportunity for building social network frameworks that support easy creation of niche-specific SNS platforms, which would support different specific contents.

Regarding privacy access control, it is significant that every SNS support public sharing. Regarding other type of settings, there is a distribution of available options, but with limited support (25\%), being significant the role of Facebook in this field.

Unary rating (like) is well supported along surveyed SNS. We cannot say the same of other types of rating, whose presence is anecdotal.

The analysis also reveals home pages as a place for publishing contents and following activity. On the other hand, profile pages are the place for content and contact lists, and less frequently activity timelines and walls.

Finally, other social entities such as groups and institutions are supported in half of the cases, not as many as their role in society might suggest.

\section{Conclusions}


Our analysis show some features being present in every SNS, i.e. user registration, relationship establishment, comments, public sharing of contents, system notifications and profile pages with avatars. User search, private messages and home pages are also very popular.

There is a strong tendency to use external services for authentication and contact recognition, despite the analyzed sites compete with each other for user base.

Relation types are almost equally distributed between unidirectional (follow) and bidirectional (friend) contacts. Contact management and private content access control is not very popular.

There is a wide set of supported content types, being text, pictures, videos, events and links the most popular ones in that order. The set is probably biased by the sample, appearing more content types if more SNS were analyzed.

The home page is primary a section to get the updates from followers and posting new things, whilst profile pages are the place to show relevant contents and contact lists.

\bibliographystyle{apalike}
\bibliography{access-control-social-networks,cscw,identity,mvc,social-networks}

\begin{thebibliography}{}

\bibitem[Barnes, 1954]{Barnes:1954}
Barnes, I. (1954).
\newblock {Class and committees in a Norwegian island parish}.
\newblock {\em Human Relations}, (7):39--58.

\bibitem[boyd and Ellison, 2007]{Boyd:2007}
boyd, d.~m. and Ellison, N.~B. (2007).
\newblock Social network sites: Definition, history, and scholarship.
\newblock {\em Journal of Computer-Mediated Communication}, 13(1):210--230.

\bibitem[King et~al., 2009]{King:2009}
King, I., Li, J., and Chan, K.~T. (2009).
\newblock A brief survey of computational approaches in social computing.
\newblock In {\em Neural Networks, 2009. IJCNN 2009. International Joint
  Conference on}, pages 1625 --1632.

\bibitem[Liduo and Yan, 2010]{Liduo:2010}
Liduo, H. and Yan, C. (2010).
\newblock Design and implementation of web content management system by
  j2ee-based three-tier architecture: Applying in maritime and shipping
  business.
\newblock In {\em Information Management and Engineering (ICIME), 2010 The 2nd
  IEEE International Conference on}, pages 513 --517.

\bibitem[McKeever, 2003]{McKeever:2003}
McKeever, S. (2003).
\newblock Understanding web content management systems: evolution, lifecycle
  and market.
\newblock {\em Industrial Management and Data Systems}, 103:686 -- 692.

\bibitem[Mislove et~al., 2007]{Mislove:2007}
Mislove, A., Marcon, M., Gummadi, K.~P., Druschel, P., and Bhattacharjee, B.
  (2007).
\newblock Measurement and analysis of online social networks.
\newblock In {\em Proceedings of the 7th ACM SIGCOMM conference on Internet
  measurement}, IMC '07, pages 29--42, New York, NY, USA. ACM.

\bibitem[Sparling and Sen, 2011]{Sparling:2011}
Sparling, E.~I. and Sen, S. (2011).
\newblock Rating: how difficult is it?
\newblock In {\em Proceedings of the fifth ACM conference on Recommender
  systems}, RecSys '11, pages 149--156, New York, NY, USA. ACM.

\bibitem[Tapiador et~al., 2011a]{Tapiador:2011:TieRBAC}
Tapiador, A., Carrera, D., and Salvach\'{u}a, J. (2011a).
\newblock Tie-rbac: An application of rbac to social networks.
\newblock {\em Web 2.0 Security and Privacy Workshop}.

\bibitem[Tapiador et~al., 2011b]{Tapiador:2011:ExtendedIdentity}
Tapiador, A., Fumero, A., and Salvachúa, J. (2011b).
\newblock Extended identity for social networks.
\newblock In Breslin, J., Burg, T., Kim, H.-G., Raftery, T., and Schmidt,
  J.-H., editors, {\em Recent Trends and Developments in Social Software},
  volume 6045 of {\em Lecture Notes in Computer Science}, pages 162--168.
  Springer Berlin / Heidelberg.
\newblock 10.1007/978-3-642-16581-8\_17.

\bibitem[Wasserman and Faust, 1994]{Wasserman:1994}
Wasserman, S. and Faust, K. (1994).
\newblock {\em {Social network analysis: methods and applications}}.
\newblock Structural analysis in the social sciences. Cambridge University
  Press.

\end{thebibliography}

\end{document}